\documentclass{sigchi-ext}
\usepackage[T1]{fontenc}
\usepackage{textcomp}
\usepackage[scaled=.92]{helvet} 
\usepackage{graphicx} 
\usepackage{balance}  
\usepackage{booktabs} 
\usepackage{ccicons}  
\usepackage{ragged2e} 

\def\plaintitle{Addressing the Privacy Implications of Mixed Reality: A Regulatory Approach} 
\def\emptyauthor{}
\def\plainkeywords{mixed reality; privacy; law}

\title{Addressing the Privacy Implications of Mixed Reality: A Regulatory Approach}

\numberofauthors{2}
\author{%
  \alignauthor{%
    \textbf{Nicole Shadowen}\\
    \affaddr{Mozilla} \\
    \email{nshadowen@mozilla.com} }\alignauthor{%
    \textbf{Diane Hosfelt}\\
    \affaddr{Mozilla}\\
    \email{dhosfelt@mozilla.com} } \vfil \alignauthor{%
    } }

\definecolor{linkColor}{RGB}{6,125,233}
\hypersetup{%
  pdftitle={\plaintitle},
  pdfauthor={\emptyauthor},
  pdfkeywords={\plainkeywords},
  bookmarksnumbered,
  pdfstartview={FitH},
  colorlinks,
  citecolor=black,
  filecolor=black,
  linkcolor=black,
  urlcolor=linkColor,
  breaklinks=true,
}

\begin{document}

\CopyrightYear{2020}
\setcopyright{rightsretained}
\conferenceinfo{CHI'20,}{April  25--30, 2020, Honolulu, HI, USA}
\isbn{978-1-4503-6819-3/20/04}
\doi{https://doi.org/10.1145/3334480.XXXXXXX}
\copyrightinfo{\acmcopyright}

\maketitle

\RaggedRight{} 

\begin{abstract}
  Mixed reality (MR) technologies are emerging into the mainstream with affordable devices like the Oculus Quest. These devices blend the physical and virtual in novel ways that blur the lines that exist in legal precedent, like those between speech and conduct. In this paper, we discuss the challenges of regulating immersive technologies, focusing on the potential for extensive data collection, and examine the trade-offs of three potential approaches to protecting data privacy in the context of mixed reality environments.
 
\end{abstract}

\keywords{\plainkeywords}

\ccsdesc[500]{Social and professional topics~Computing / technology policy}
\ccsdesc[300]{Security and privacy~Privacy protections}
\ccsdesc[100]{Human-centered computing~Human computer interaction (HCI)}

\section{Introduction}
MR devices offer businesses and consumers a landscape of opportunity for deeper interaction, immersion, image projection, and information. Using both the digital and real world, mixed reality technologies like virtual reality (VR) and augmented reality (AR), occlude the users’ senses to project an experience that feels real. 

As with most technology, regulation lags innovation and faces many challenges. Too often, we only protect data after a large breach has occurred. In MR, large amounts of data are biometrics or biometrically-derived data (BDD), which, unlike passwords or credit card numbers, are tied to our intrinsic characteristics and cannot be changed after a breach. Therefore, it's imperative to require protection before a breach happens.

In this paper, we build on our exploration of the privacy implications of eye tracking and other biometric data use in these technologies, by contextualizing the risks through current regulatory approaches~\cite{hosfeltshad2020}. These are not meant to be prescriptive, but instead to inspire debate about how best to manage the inherent risk of large amounts of biometric and other sensitive data types incurred by MR systems.

First, we will briefly discuss the challenges to regulating immersive technologies, where a user's senses, motion, and body characteristics, as well as information about the world around them, may be recorded and used to provide a high fidelity experience and account for optimization. Our focus is regulating the data output from an MR experience. Then we will discuss three different potential approaches for regulating this data (focused on U.S. law), contextualized through related research in the human-computer interaction (HCI) community. This work focuses on legal and regulatory approaches to protect privacy related to biometric and data identity in MR, as listed in Bye’s MR Ethics framework~\cite{Bye}.


In immersive MR, a user’s information is required to make the experience work.  Sensors and cameras track the body’s position and orientation in space, movements, characteristics, and responses to a variety of sensory stimuli. This provides the input needed to produce a compelling environment within which the user may interact, while also giving developers feedback on features that don’t work, or even make consumers feel dizzy or ill~\cite{lemley2017law} (giving a new meaning to software bugs). This information is mostly comprised of biometrically-derived data, BDD, a category of data that is derived from a user's intrinsic traits, characteristics, and biometrics. For example, while a retina scan is considered biometric data, the data output from eye tracking would be considered BDD~\cite{Hosfelt}.  Typically, this data is recorded on the user’s device and in the cloud.

\section{Challenges to regulation}
Despite the apparent risk of unbridled access to sensitive user data mixed reality systems may pose, legislative movement has been slow. One explanation for this may be existing law and normative precedent in the technology provider to consumer relationship. In the 1996 Communications Decency Act (CDA), section 47 U.S.C. 230, the law states, "no provider or user of an interactive computer service shall be treated as the publisher or speaker of any information provided by another information content provider"~\cite{effcda}. In many cases, this provision has served to shield technology providers from the content that users create and publish on their platforms. In an immersive environment, this may extend to include the content and behaviors of users in the VR space---blurring the lines of liability when it comes to how data is shared between users and with providers.


Beyond norms that have deterred change are philosophical and fundamental questions that make legislating mixed reality data collection and use exceedingly complex. These include the blurred lines between speech and conduct as well as physical and psychological harm.  In the tangible world, the line between physical conduct and verbal speech is easier to distinguish than that of a virtual world, because in a virtual environment, most behavior and experience is simulated and not truly physical. Despite the lack of touch, these experiences still feel real to users.  

Jordan Belamire experienced virtual sexual harassment just as commercial VR was becoming available. “The virtual groping,” she said, “feels just as real. Of course, you’re not physically being touched, just like you’re not actually one hundred feet off the ground, but it’s still scary as hell"~\cite{belamireVR}.

If all behavior, action, and speech in immersive MR is considered “speech” because it is not completely tangible, it may be protected and go un-penalized under the First Amendment~\cite{firstamendment}. Similarly, psychological harm, beyond physical harm, will need further understanding and protective legislation in these spaces. As MR creators and providers interact with behavioral and generated data from their ecosystems, regulatory requirements will be needed to parse the data types and how they may be handled based on the risks to users. 

\section {Discussion of regulatory approaches}
Through their very use, immersive MR experiences incur large amounts of sensitive user data.  Based on the characteristics of this data, many users may want more say over who has access to it, and legislative bodies may be more motivated to regulate due to a higher risk of user vulnerability.  In identifying the best ways to regulate this type of information, it is useful to understand the ethical trade-offs of approaches through the context of both technology driven methodologies and current U.S. law. Using this, we can find paths forward that marry regulatory and technological solutions in an HCI context to solve for privacy concerns in MR.

\subsection{Code as law}
When developers of QuiVr were made aware of the virtual groping that occurred in their immersive world to user, Jordan Belamire, they acted quickly to find ways to solve the problem. Ultimately, the game’s engineers adapted code to extend a “personal bubble” around users’ entire bodies if another user reached for them \cite{belamireVR}.  In doing so, they provided users with the agency to protect themselves from perceived aggressors in the environment, effectively “outlawing” unwanted groping and other related incidents.

Similar to this in-world example, “code is law”, a concept first popularized by Lawrence Lessig in 1999, references the ability of technology builders and creators to mandate “regulations” into their users’ experiences by writing preferred features, behaviors, and scenarios into the code, and leaving others out.  In this way, immersive MR systems can be designed with privacy in mind - from interaction features, like the personal bubble in QuiVR, to system architecture, such as, local data processing for privacy preservation. 

\subsection{BDD as health data}
Under HIPAA rules, individuals have the right to obtain a copy of their protected health information (PHI), request corrections and limits, as well as know the identity of who has received the records \cite{oversight2016}. Typically, PHI under HIPAA encompasses a patient’s medical record or payment history, and is considered most sensitive when linked to a list of eighteen protected identifiers: ranging from name and phone number to biometric identifiers like finger and voice prints \cite{hipaajournal}. 

While HIPAA sets a precedent for how to handle data over which an individual may have limited control and which may make them vulnerable to scrutiny or discrimination, HIPAA coverage is applied to entities not to types of data. Therefore, to fit into the pre-existing structure, HIPAA could not be invoked for specific categories of data, such as health or biometric, but for commercial entities deemed to be interacting with them. For example, this protects the scenario in which Alice confides PHI to Bob, who then accidentally tells Eve and Mallory without Alice's consent from legal action.

This brings into question when this decision would be made, by which criteria, and by whom. The Department of Health and Human Services, has resisted extending HIPAA to entities beyond health plans, health care clearing houses, and health care providers. However, noticing gaps in these types of oversight, policymakers have used other methods of regulating bad data use practices, such as through the FTC Act to investigate “unfair or deceptive practices” \cite{ftcact}\cite{oversight2016}. Despite the drawbacks, using HIPAA as a framework for understanding the best basic protections for users’ sensitive data, including biometrics, could be an important way to build on current regulatory practices.

\subsection{Data as a property right}
Data as a property right proposes to flip the current paradigm on its head, putting consumers in control of their data and able to use it as a commodity in the same ways technology developers use their users’ data today. This concept was recently espoused by U.S. presidential candidate, Andrew Yang, as a part of his proposed policy to regulate technology firms in the 21st century \cite{yang2020}. In Yang’s version of the right to data ownership, citizens should be guaranteed the right to know who will use their data, opt-out of collection and sharing, have transparency about ownership changes and data breaches, as well as full data portability.  These mandates would affect a wide variety of technology creators, including those creating immersive MR spaces.  In order to comply with these requirements, immersive experiences may be completely altered.  

Data as a property right also poses problems for the users it claims to protect. Taking into account the ambivalence with which many individuals click through privacy terms and conditions, owning or controlling one’s data may not be most people’s first priority or interest \cite{royalsociety2018}. Last, treating data as a property right, or commodity, may constrain the way people think about or prioritize their privacy.  Grossklags found strong support for the hypothesis that people may be willing to sell their data for money. In fact, most participants in their empirical study were willing to sell their personal data for little more than 25 cents, and many waived protection of their data in the course of doing so. \cite{grossklags25}. Treating data as property, which users can own and trade, could alter the criteria by which decisions about data are made, reducing the principle of privacy as a right to commodity as a trade. This shift may have an unprecedented and unexpected impact on the layers of society, and requires further understanding, both technically and philosophically before implementation.

\section{Conclusion}
Immersive mixed reality promises to bring consumers and stakeholders a range of delightful, meaningful, and compelling experiences beyond the realm of possibility.  However, what is required to build these experiences may be more than our society or legal system has seen to date. Using a genuine understanding of the experiences to be had, as well as legal, technical, and social approaches that have come before, we can build these systems consciously with respect for individual dignity and privacy.

\balance{} 

\bibliographystyle{SIGCHI-Reference-Format}
\bibliography{bib}

\end{document}